# Comment on "generalized Gibbs' approach"
# [J. Chem. Phys. 119, 6166 (2003); 124, 194503 (2006)]

Nikolay V. Alekseechkin

Akhiezer Institute for Theoretical Physics, National Science Centre "Kharkov Institute of Physics and Technology", Akademicheskaya Street 1, Kharkov 61108, Ukraine

E-mail: n.alex@kipt.kharkov.ua

"Generalized Gibbs' approach" of Schmelzer et al [J. Chem. Phys. 119, 6166 (2003); 124, 194503 (2006)] results in wrong conditions of equilibrium of a critical embryo with the ambient phase. The presented analysis shows the shortcomings of this approach and proves the thermodynamic conditions of equilibrium. Also, a generalized adsorption equation is obtained in different forms.

In Refs. 1, 2, as well as in many related articles of Schmelzer et al, the so called "generalized Gibbs' approach" is proposed for a nonequilibrium treatment of the cluster formation. The impressive result of this approach is the conditions of equilibrium of a critical embryo with the ambient phase[1,2]

$$(T_\beta - T_\alpha) = \frac{3}{R_*}\left(\frac{\partial \sigma}{\partial s_\alpha}\right)_{\{\phi_\beta\}} \tag{1a}$$

$$(\mu_{i\beta} - \mu_{i\alpha}) = \frac{3}{R_*}\left(\frac{\partial \sigma}{\partial \rho_{i\alpha}}\right)_{\{\phi_\beta\}}, \quad i = 1, 2, \ldots k, \tag{1b}$$

where the indices $\alpha$ and $\beta$ relate to new and old phases, respectively; $R_*$ is the critical radius, $\rho_{i\alpha}$ and $s_\alpha$ are the volume densities of the $i$-species particles and entropy in the cluster; $\sigma$ is the surface tension; $\{\phi_\beta\}$ is the set of intensive parameters describing the ambient phase state.

In other words, the temperatures, as well as the chemical potentials, of two parts of a system in equilibrium are not equal. As is known, the inequality of these quantities results in the flows of heat and substance between a cluster and the ambient phase (only if the cluster is not isolated from the latter). However, the presence of these flows is incompatible with equilibrium. Some physical consequences of these "conditions of equilibrium" are discussed at the end.

The reason of such "modification" of thermodynamic conditions of equilibrium is attributed by the authors of Refs. 1, 2 to the dependence of the surface tension on embryo state parameters. According to this fact, the authors start with the generalization of the fundamental Gibbs equation for incrementing the superficial energy $U_\sigma$,

$$dU_\sigma = T_\beta dS_\sigma + \sum_{i=1}^{k} \mu_{i\beta} dn_{i\sigma} + \sigma dA + \sum_{i=1}^{k+1} \Phi_{i\alpha} d\phi_{i\alpha},  \quad (2)$$

where $S_\sigma$ and $n_{i\sigma}$ are the superficial entropy and the superficial number of particles[4] of the $i$-th component; $\phi_{i\alpha}$ is the set of intensive parameters describing the state of a cluster:

$$\phi_{i\alpha} = \rho_{i\alpha} \text{ for } i = 1, 2, \ldots, k; \quad \phi_{k+1} = s_\alpha  \quad (3)$$

Further, applying the procedure leading to Gibbs' adsorption equation, the authors come to "generalized Gibbs' adsorption equation"

$$S_\sigma dT_\beta + Ad\sigma + \sum_{i=1}^{k} n_{i\sigma} d\mu_{i\beta} = \sum_{i=1}^{k+1} \Phi_{i\alpha} d\phi_{i\alpha}  \quad (4)$$

While the general form of Eq. (4) is not objectionable in view of the postulated dependence $\sigma(\phi_{i\alpha})$, the form of Eq. (2) and the choice of the quantities $\phi_{i\alpha}$, Eqs. (3), are questionable. First, the form of the last term in Eq. (2) (which is the same as in Eq. (4)) is not substantiated, it is simply postulated. It is not clear why e.g. the term $\sum_{i=1}^{k+1} \phi_{i\alpha} d\Phi_{i\alpha}$ is not employed instead of that. Second, the choice of the quantities $\phi_{i\alpha}$ is strange; it is not clear why the same type of the dependence of the surface tension as in the original Gibbs adsorption equation – on temperatures and chemical potentials – does not hold in the generalized adsorption equation (especially in view of the fact that the latter has to be reduced to the former in equilibrium). The set $\{T_\alpha, P_\alpha, \{x_{i\alpha}\}\}$, $i = 1, 2, \ldots, k-1$, where $x_{i\alpha}$ is the molar fraction of the $i$-th component in a cluster and $P_\alpha$ is the pressure, is shown below as an appropriate set for the cluster state description. At the same time, Eqs. (1) are derived in Ref. 1 from the condition $dU_{tot} = 0$, where $U_{tot}$ is the energy of the two-phase system "cluster + ambient phase"; i.e. they are a consequence of Eqs. (2) and (3).

In order to get the conditions of equilibrium of a cluster with the ambient phase, one needs to start with a proper generalization of the equilibrium equation for the energy $U_{tot}$ to a nonequilibrium case (i.e. to get an equation for the energy of the system "noncritical cluster + ambient phase"). Such analysis is carried out in Ref. 3; here it is generalized to a multicomponent case. The key idea in this analysis is to split the superficial entropy $S_\sigma$ and the superficial number of particles $n_{i\sigma}$ into two parts,

$$S_\sigma = S_{\sigma\alpha} + S_{\sigma\beta}, \quad n_{i\sigma} = n_{i\sigma\alpha} + n_{i\sigma\beta},  \quad (5)$$

where $S_{\sigma\alpha}$ and $S_{\sigma\beta}$ are the contributions to $S_\sigma$ from the phases $\alpha$ and $\beta$, respectively; the same holds for $n_{i\sigma\alpha}$ and $n_{i\sigma\beta}$. The quantities $S_{\sigma\alpha}$, $S_{\sigma\beta}$, $n_{i\sigma\alpha}$, and $n_{i\sigma\beta}$ can be called the "one-sided" superficial quantities; these are the corrections to the bulk values of entropy and numbers of particles in order to obtain the true values of these quantities for the phases $\alpha$ and $\beta$. Considering the thermodynamics of a surface layer, Gibbs represents this layer as consisting of two parts with volumes $\upsilon'''$ and $\upsilon''''$ separated by the dividing surface[4]. The bulk entropies of these parts are denoted in Ref. 4 as $\eta'''$ and $\eta''''$; $\eta$ is the true entropy of the layer. In accordance with Eq. (5),

$$\eta = \eta''' + \eta'''' + S_\sigma = (\eta''' + S_{\sigma\alpha}) + (\eta'''' + S_{\sigma\beta}) \qquad (6)$$

The similar relation also takes place for the masses $m_i'''$ and $m_i''''$ or the numbers of particles. So, the separation given by Eqs. (5) is fully consistent with Gibbs' approach.

The energy $U_{tot}$, in view of Eqs. (5), naturally splits into two parts – the cluster energy $U$ and the ambient phase energy $U_0$:

$$U_{tot} = U + U_0, \qquad (7)$$

where[3]

$$U = T_\alpha S - P_\alpha V + \sum_{i=1}^{k} \mu_{i\alpha} n_i + \sigma A \qquad (8a)$$

$$U_0 = T_\beta S_0 - P_\beta V_0 + \sum_{i=1}^{k} \mu_{i\beta} n_{i0} \qquad (8b)$$

and

$$S = S_\alpha + S_{\sigma\alpha}, \quad n_i = n_{i\alpha} + n_{i\sigma\alpha}, \quad S_0 = S_\beta + S_{\sigma\beta}, \quad n_{i0} = n_{i\beta} + n_{i\sigma\beta} \qquad (9)$$

Here $V$ and $V_0$ are the volumes of the cluster and the ambient phase, respectively; $S_\alpha$, $S_\beta$, $n_{i\alpha}$, and $n_{i\beta}$ are the bulk values of entropy and numbers of particles in each of the phases; $S$, $S_0$, $n_i$, and $n_{i0}$ are the *true* values of entropy and numbers of particles in each of the phases; $A$ is the cluster surface area.

In the state of equilibrium, $T_\alpha = T_\beta$ and $\mu_{i\alpha} = \mu_{i\beta}$, the equation for $U_{tot}$ takes the known equilibrium form[5]

$$U_{tot} = T_\beta S_{tot} - P_\alpha V - P_\beta V_0 + \sigma A + \sum_{i=1}^{k} \mu_{i\beta} n_{i,tot}, \qquad (10)$$

where $S_{tot} = S + S_0$, $n_{i,tot} = n_i + n_{i0}$, and $V_{tot} = V + V_0$.

The superficial energy $U_\sigma$ is distributed between the quantities $U$ and $U_0$, in view of Eqs. (9); it is expressed via the one-sided superficial quantities as follows[3]:

$$U_\sigma = T_\alpha S_{\sigma\alpha} + T_\beta S_{\sigma\beta} + \sigma A + \sum_{i=1}^{k} \mu_{i\alpha} n_{i\sigma\alpha} + \sum_{i=1}^{k} \mu_{i\beta} n_{i\sigma\beta}$$

$$= U_\sigma^{(eq)} + (T_\alpha - T_\beta)S_{\sigma\alpha} + \sum_{i=1}^{k}(\mu_{i\alpha} - \mu_{i\beta})n_{i\sigma\alpha}, \tag{11a}$$

$$U_\sigma^{(eq)} = T_\beta S_\sigma + \sigma A + \sum_{i=1}^{k} \mu_{i\beta} n_{i\sigma}, \tag{11b}$$

where the equation for $U_\sigma^{(eq)}$ has the same form as for the equilibrium value of $U_\sigma$ in Gibbs' approach.

The equation for $dU_\sigma$, as well as a generalized adsorption equation (for a noncritical cluster which is not in equilibrium with the ambient phase), acquire the following form[3]:

$$dU_\sigma = T_\beta dS_\sigma + \sum_{i=1}^{k}\mu_{i\beta}dn_{i\sigma} + \sigma dA + (T_\alpha - T_\beta)dS_{\sigma\alpha} + \sum_{i=1}^{k}(\mu_{i\alpha} - \mu_{i\beta})dn_{i\sigma\alpha} \tag{12}$$

$$Ad\sigma = -S_\sigma dT_\beta - \sum_{i=1}^{k} n_{i\sigma} d\mu_{i\beta} - S_{\sigma\alpha}d(T_\alpha - T_\beta) - \sum_{i=1}^{k} n_{i\sigma\alpha} d(\mu_{i\alpha} - \mu_{i\beta}) \tag{13}$$

Eq. (13) determines the dependence of the surface tension on state parameters of both the phases. If we are interested in the dependence on cluster state parameters only, we have to fix the state of the ambient phase, i.e. to put $T_\beta$ and $\mu_{i\beta}$ constant (such situation corresponds to nucleation in a large reservoir and often occurs in practice); Eq. (13) becomes

$$(Ad\sigma)_{T_\beta, \mu_{i\beta}} = -S_{\sigma\alpha} dT_\alpha - \sum_{i=1}^{k} n_{i\sigma\alpha} d\mu_{i\alpha} \tag{14}$$

Being combined with the Gibbs-Duhem equation for the bulk phase $\alpha$,

$$-S_\alpha dT_\alpha - \sum_{i=1}^{k} n_{i\alpha} d\mu_{i\alpha} + V dP_\alpha = 0, \tag{15}$$

it reads as

$$(Ad\sigma)_{T_\beta, \mu_{i\beta}} = -SdT_\alpha + VdP_\alpha - \sum_{i=1}^{k} n_i d\mu_{i\alpha} \tag{16a}$$

If the state of the ambient phase is not assumed to be fixed, then the following equation is derived in the same way:

$$Ad\sigma = \left[-SdT_\alpha + VdP_\alpha - \sum_{i=1}^{k} n_i d\mu_{i\alpha}\right] + \left[-S_0 dT_\beta + V_0 dP_\beta - \sum_{i=1}^{k} n_{i0} d\mu_{i\beta}\right] \tag{16b}$$

Again, the known equilibrium equation[5] $Ad\sigma = -S_{tot}dT_\beta + VdP_\alpha + V_0 dP_\beta - \sum_{i=1}^{k} n_{i,tot} d\mu_{i\beta}$ follows from Eq. (16b).

The chemical potential $\mu_{i\alpha}$ is a function of $T_\alpha$, $P_\alpha$, and the composition $\{x_{i\alpha}\}$. Eqs. (14) and (16) determine the dependence of the surface tension on these cluster state parameters. For this reason, the set $\{T_\alpha, P_\alpha, \{x_{i\alpha}\}\}$ was noted above as an appropriate one.

Comparing Eq. (12) with Eq. (2), we find the following correspondence:

$$\sum_{i=1}^{k} \Phi_{i\alpha} d\phi_{i\alpha} \to (T_\alpha - T_\beta) dS_{\sigma\alpha} + \sum_{i=1}^{k} (\mu_{i\alpha} - \mu_{i\beta}) dn_{i\sigma\alpha} \qquad (17)$$

At the same time, the comparison of Eqs. (13) and (4) yields

$$\sum_{i=1}^{k} \Phi_{i\alpha} d\phi_{i\alpha} \to -S_{\sigma\alpha} d(T_\alpha - T_\beta) - \sum_{i=1}^{k} n_{i\sigma\alpha} d(\mu_{i\alpha} - \mu_{i\beta}) \qquad (18)$$

The discrepancy between Eqs. (17) and (18) shows that Eqs. (2) and (4) cannot contain the same term $\sum_{i=1}^{k} \Phi_{i\alpha} d\phi_{i\alpha}$. Since the quantities $\phi_{i\alpha}$ are intensive, we have to employ Eq. (18) for their determination. Thus, one obtains

$$\phi_{i\alpha} = \mu_{i\alpha} - \mu_{i\beta} \text{ for } i = 1, 2, \ldots, k \,;\, \phi_{(k+1)\alpha} = T_\alpha - T_\beta, \qquad (19)$$

which is quite different from the set given by Eqs. (3). We see that the quantities $\phi_{i\alpha}$ depend on the states of both the cluster and ambient phase; they describe the cluster state only if the ambient phase state is fixed. In the equilibrium state, $\phi_{i\alpha} = 0$ and Eq. (13) converts to Gibbs' adsorption equation, as it must. Also, Eqs. (11a) and (12) have their usual, equilibrium, form. Thus, the distinction of all these equations from the corresponding Gibbs equations is due to the nonequilibrium sate of the considered system, $T_\alpha \neq T_\beta$ and $\mu_{i\alpha} \neq \mu_{i\beta}$. One more conclusion from Eqs. (17) and (18) is that Eq. (2) has to contain the term $-\sum_{i=1}^{k} \phi_{i\alpha} d\Phi_{i\alpha}$, as was noted above, with $\Phi_{i\alpha} = -n_{i\sigma\alpha}$ for $i = 1, 2, \ldots, k$ and $\Phi_{(k+1)\alpha} = -S_{\sigma\alpha}$.

In order to derive the conditions of equilibrium, we consider a cluster as a body in a medium[6] or a system in a thermostat[7]. For such body, the quantity

$$\left[ U - T_\beta S + P_\beta V - \sum_{i=1}^{k} \mu_{i\beta} n_i \right]$$

has a minimum in the state of equilibrium[6,7]. This fact means

$$\delta U - T_\beta \delta S + P_\beta \delta V - \sum_{i=1}^{k} \mu_{i\beta} \delta n_i = 0 \qquad (20)$$

From Eq. (8a),

$$\delta U = \left[ T_\alpha \delta S - P_\alpha \delta V + \sum_{i=1}^{k} \mu_{i\alpha} \delta n_i + \sigma \delta A \right] + \left[ S \delta T_\alpha - V \delta P_\alpha + \sum_{i=1}^{k} n_i \delta \mu_{i\alpha} + A \delta \sigma \right] \qquad (21)$$

This equation is the full variation of $U$; the expression in the second brackets takes into account the possible fluctuations of intensive parameters of the cluster. However, this expression is equal to zero, in view of Eq. (16a). Substituting Eq. (21) into Eq. (20), we get

$$(T_\alpha - T_\beta)\delta S - (P_\alpha - P_\beta)\delta V + \sigma\delta A + \sum_{i=1}^{k}(\mu_{i\alpha} - \mu_{i\beta})\delta n_i = 0,$$

from where the well-known thermodynamic conditions of equilibrium[4-10] follow:

$$T_\alpha = T_\beta, \quad \mu_{i\alpha} = \mu_{i\beta}, \quad P_\alpha - P_\beta = 2\sigma/R_* \qquad (22)$$

As is seen, the question of the dependence of the surface tension on cluster state parameters arising in Eq. (21) in the second brackets is simply resolved with the help of Eq. (16a): this dependence does not change the fundamental equation for $dU$ and the thermodynamic conditions of equilibrium.

The second way to derive Eqs. (22) is to use general Gibbs' condition of equilibrium[4]

$$(dU_{tot})_{S_{tot}, V_{tot}, n_{i,tot}} = 0 \qquad (23)$$

Employing Eqs. (8), we find

$$dU_{tot} = d(U + U_0) = T_\beta dS_{tot} - P_\beta dV_{tot} + \sum_{i=1}^{k}\mu_{i\beta}dn_{i,tot} + \left[SdT_\alpha - VdP_\alpha + \sum_{i=1}^{k}n_i d\mu_{i\alpha} + Ad\sigma\right]$$

$$+ \left[S_0 dT_\beta - V_0 dP_\beta + \sum_{i=1}^{k}n_{i0}d\mu_{i\beta}\right] + (T_\alpha - T_\beta)dS - (P_\alpha - P_\beta)dV + \sigma dA + \sum_{i=}^{k}(\mu_{i\alpha} - \mu_{i\beta})dn_i \quad (24)$$

Here the ambient phase state is not fixed, so its parameters are also varied. From Eqs. (23), (24), and (16b), Eqs. (22) follow also.

Finally, the third way to derive Eqs. (22) is to employ the extremum condition $dW = 0$ for the cluster formation work $W$ derived in Ref. 3 at a fixed state of the ambient phase:

$$W = \sum_{i=1}^{k}(\mu_{i\alpha} - \mu_{i\beta})n_i + (T_\alpha - T_\beta)S - (P_\alpha - P_\beta)V + \sigma A \qquad (25)$$

The equation $dW = 0$ is also the condition of equilibrium as a consequence of the extremum properties of entropy[3]; together with Eq. (16a), it leads to Eqs. (22).

The appeal of the authors to a certain similarity of Eqs. (1) and Eq. (22) for pressures is not valid. Pressure is both a mechanical and thermodynamic quantity. A two-dimensional stretched film is the mechanical equivalent of the surface layer, when the surface of tension is employed as a separating surface[5]. Thus, the Laplace pressure, $2\sigma/R_*$, is a real physical quantity – the pressure produced by this film. Eq. (22) for pressures is simply the condition of mechanical equilibrium; it is obtained both in mechanics and thermodynamics. As regards the temperature, this is exclusively thermodynamic quantity; it has no a mechanical analogue. So, the mentioned comparison is incorrect. It is more relevant to compare Eq. (1a) with Newton's law of heat exchange $j = \alpha(T_\beta - T_\alpha)$, where $j$ is the heat flux density, $\alpha$ is the heat transfer coefficient. Calling Eq. (1a) the "equilibrium condition" (which implies $j = 0$), the authors reject this law.

The statement of the authors[1,2] that Eqs. (22) for temperature and chemical potentials are applicable to a planar interface only (i.e. at $R \to \infty$) is wrong. Gibbs[4] studied just curved interfaces, so that the equilibrium conditions (22) where derived by him just for this case. Furthermore, he discussed the limit $R \to 0$ (when the surface of tension vanishes, but some inhomogeneity remains[5]) as well.

It is seen that the differences in Eqs. (1) increase with decreasing $R_*$ infinitely at $R_* \to 0$. Thus, depending on the sign of the right side of Eq. (1a), we have either "strongly superheated droplets" in a metastable vapor (without the heat exchange with the vapor!) vs. "strongly supercooled bubbles" in a metastable liquid or vice versa ("crystal particles" in the vapor vs. "superheated bubbles" in the liquid). The authors do not estimate the maximum degree of superheating with the corresponding value of $R_* = R_{*,\lim}$ and do not answer the question how the embryos with $R_* < R_{*,\lim}$ nucleate. The case of a "supercooled embryo" is even more interesting: the embryo temperature $T_\alpha$ cannot decrease infinitely, it is limited to the value $T_\alpha = 0$. The authors do not estimate the value of $R_*$ at which the absolute zero is achieved. Anyway, the authors have found a simple way of getting ultralow temperatures: it is enough to implement a nucleation experiment properly!

Let us assume that we have the case of "strongly superheated droplets". Firstly, the nucleation of real crystal embryos at sufficiently low temperatures of vapor is most likely impossible (they are "superheated" and therefore liquid). Secondly, the arising question is: whence droplets receive the heat and in what way? They could receive it via the heat exchange either from an external source with $T > T_\alpha$ or from the mother phase with the temperature $T_\beta < T_\alpha$ contrary to the second law of thermodynamics. A critical droplet is in unstable equilibrium with the mother phase. When it grows, the system evolves to the stable equilibrium state. Does a "superheated droplet" give the excessive energy (of unknown origin) to the mother phase in this process? As is known, the existence of a *perpetuum mobile* is forbidden by the 1st and 2nd laws of thermodynamics. The thermodynamic conditions of equilibrium, Eqs. (22), are a consequence of these laws (as is seen from above derivations), whereas Eqs. (1) are incompatible with them. So, the "thermodynamics" based on the "conditions of equilibrium" (1) enables a perpetuum mobile.

The presented analysis results in the following conclusions.

(i) The dependence of the surface tension on cluster state parameters does not change the thermodynamic conditions of equilibrium, as it must from the physical point of view. The conditions of equilibrium are a consequence of the fundamentals of thermodynamics (the first and the second laws)[4-10] and therefore they are quite general; they cannot depend on specific properties of any thermodynamic quantity, such as the dependence of $\sigma$ on cluster state parameters. Gibbs wrote[4]: "… the surfaces of discontinuity do not affect the equilibrium conditions related to temperature and [chemical] potentials". So, the approach of Refs. 1, 2 leads to wrong conditions of equilibrium.

(ii) The stationary nucleation rate (which is of the most practical interest) is determined by the parameters of a critical cluster, i.e. by equilibrium properties of the system "cluster + ambient phase" including the equilibrium value $\sigma_*$ of the surface tension.[3] Thermodynamic parameters of a critical cluster are determined by Eqs. (22). At sufficiently large values of $R_*$, the surface tension does not depend on $R$ in these equations ($\sigma = \sigma_*$). In the case of small values of $R_*$, the system of Eqs. (22) has to be complemented by the dependence $\sigma(R)$ the determination of which is a separate thermodynamic problem[4,5]. Since the chemical potential is a function of temperature, pressure, and composition, Eqs. (22) determine the dependences $P_\alpha(R_*)$ (or $P_\beta(R_*)$), $x_{i\alpha}(R_*)$ (the dependence of composition of a critical cluster on its radius) and the critical radius $R_*$ itself[3] (the method for deriving these dependences and resulting equations can be found in Ref. 5). So, the classical Gibbs theory is quite sufficient for calculating the parameters of a critical cluster (including the work of formation) and does not require generalizations for that.

(iii) The generalized adsorption equation has the form yielded by Eqs. (13) and (16b), or (14) and (16a); it involves the one-sided superficial quantities.